# HOLE SUPERCONDUCTIVITY FROM KINETIC ENERGY GAIN

J.E. Hirsch

Department of Physics, University of California, San Diego, La Jolla, CA 92093-0319, USA

The apparently unrelated experimental observations of optical sum rule violation[1] and of tunneling asymmetry in NIS tunneling[2] find a simple explanation within the theory of hole superconductivity. In fact, both phenomena were predicted by the theory long before they were experimentally observed[3,4]. Other experimental predictions of the theory, in particular a novel feature expected in photoemission experiments, are discussed.

Consider the motion of cars in the parking garage[5] shown in the figure to the right. To bring a single car (a) to the exit at the top is much easier than to do it for a single car 'hole' (b): in the latter case, several cars have to be moved, which may bump into other cars in the process creating a large disruption. When moving two cars, as in (c), it is wise to keep them far apart to avoid collisions; instead, when moving two car 'holes' as in (d), it is advantageous to bring them together, as this makes their propagation easier than if they move separately.

The theory of hole superconductivity postulates that electrons in energy bands in solids behave as the cars in this garage. In an almost full band, holes have difficulty propagating because they cause a large disruption in their environment when doing so. By increasing the hole concentration through doping, or by pairing which increases the 'local' hole concentration, propagation becomes easier.

This physics is represented by a decrease in the hole effective mass, or equivalently an increase in its hopping amplitude, as the local hole concentration increases. For a hole of spin $\sigma$ hopping between sites i and j, the hopping amplitude is given by

$$t_{ij}^\sigma = t_h + \Delta t (n_{i,-\sigma} + n_{j,-\sigma}) \quad (1)$$

with $n_{i,-\sigma}$ the occupation number for a hole of spin $-\sigma$ at site i, and $t_h$ and $\Delta t$ of the same sign. Eq. (1) implies that the hopping amplitude for an isolated hole, $t_h$, increases to $t_h + \Delta t$ when another hole exists at either of the two sites involved in the hopping process, as shown in Fig. 2. It can be justified from first principles calculations on small molecules[6], which show that $\Delta t$ will be largest for hole hopping between negatively charged anions, such as the $O^=$ ions in the Cu-O planes. $\Delta t$ leads to pairing of holes due to lowering of the system's kinetic energy.

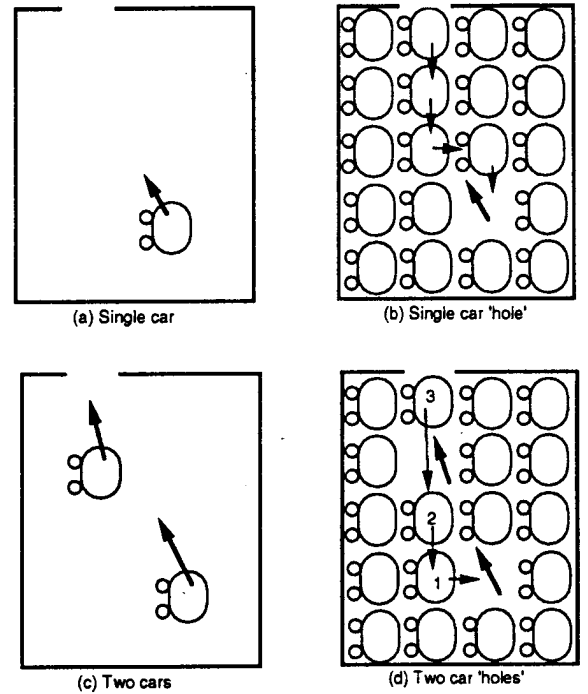

Fig. 1: Would you rather park your car in a garage that's almost empty or in one that's almost full? Where will you get it out faster? To move the two car 'holes' in (d) it is advantageous to follow the sequence shown: after moving 1 and 2 the holes are 'paired' and will move more easily.

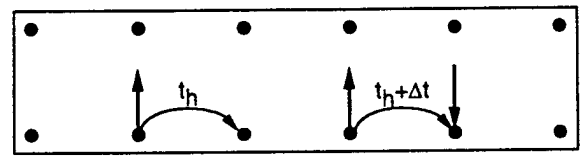

Fig. 2: The hopping amplitude for a hole is larger when another hole is nearby.

The enhanced hopping amplitude and kinetic energy lowering that occurs when carriers pair has a direct observable consequence in optical experiments, shown schematically in Fig. 3. Basov and coworkers[1] have recently found that the low frequency missing area ($\delta A_l$ in the figure) in the frequency-dependent optical conductivity can account for as little as 50% of the measured zero-frequency $\delta$-function weight in some cases. Where is the extra weight coming from? Fugol and coworkers[7] provide us with the answer, they had already earlier detected a decrease in optical absorption in frequencies in the visible range as the systems go superconducting. This rather unexpected behavior violates what had been a basic tenet of superconductivity for over 30 years, the Ferrell-Glover-Tinkham sum rule[8]. It was predicted in detail, including its temperature and doping concentration dependence (not all of it has been experimentally verified yet) in 1992[3], before any experimental evidence for it existed.

Why is lowering of kinetic energy tied to hole, rather than electron, superconductivity? We can get some understanding from consideration of the nature of electronic energy states in a band. If superconductivity is to occur through lowering of kinetic, as opposed to potential, energy, it is natural to expect that this will happen when the kinetic energy is particularly high in the normal state. This will be the case when electrons at the Fermi level are in antibonding states, shown schematically in Fig. 4. This argument illustrates that there is a natural link between carriers having low mobility in the unpaired state and increasing their mobility as well as decreasing their kinetic energy in the paired state, and their hole-like character.

There is more to it however. The reason that single hole carriers have low mobility in the normal state is that they are heavily dressed by interactions with their non-rigid electronic background (e.g. cars bumping against each other in the garage analogy). As the local hole concentration increases, either by doping or by pairing, carriers become increasingly undressed, as shown schematically in Fig. 5, and a transfer of optical spectral weight from high to low frequencies occurs. The transfer of spectral weight with hole doping in the normal state has also been seen experimentally[9]. The physical origin of the high frequency missing area $\delta A_h$ is not contained in Eq. (1), but is also explained by the theory of hole superconductivity[10]: it can be understood as a direct consequence of electron-hole asymmetry.

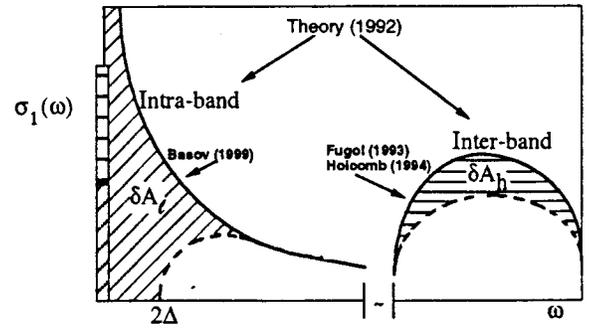

Fig. 3: Optical conductivity versus frequency (schematic) in the normal (solid) and superconducting (dashed) states. The zero-frequency $\delta$-function, that determines the penetration depth, has contributions from missing areas at low frequencies ($\delta A_l$, diagonally hatched) and high frequencies ($\delta A_h$, horizontally hatched). $\delta A_h$ is proportional to the kinetic energy lowering and is zero in a conventional BCS superconductor.

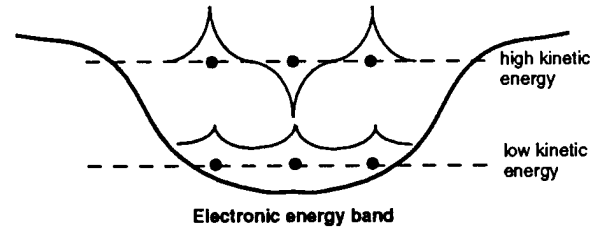

Fig. 4: Electronic energy states in a solid (schematic). The states near the top of the band (hole states) have higher kinetic energy than those at the bottom.

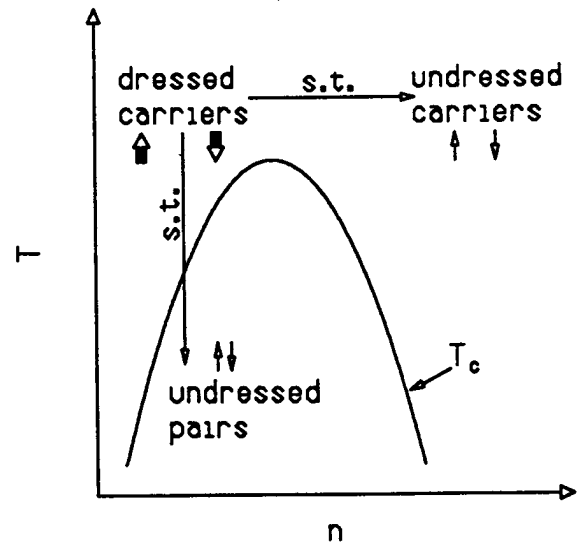

Fig. 5: Dilute hole carriers above $T_c$ have a large effective mass. As their local density increases, either by doping or by pairing, optical spectral weight is transferred from high to low frequencies and their effective mass decreases.

Is there an equally clear experimental signature of the essential underlying electron-hole asymmetry? In fact there is. Experimental NIS tunneling spectra show a consistent asymmetry in the conductance peaks, with larger current corresponding to a negatively biased sample[2]. The fact that the peak asymmetry is larger than that of the background indicates that it is an intrinsic superconducting effect, due to a gap slope, rather than due to normal state density of states or barrier effects[11]. Fig. 6 shows calculated spectra, with and without gap slope, for gap $\Delta$=30meV, $T/T_c$=0.1, and broadening parameter $\Gamma$=3meV. The asymmetry was predicted in 1989[4], before experimental evidence for it existed.

This effect is a direct consequence of Eq. (1), which leads to a linear dependence of the superconducting energy gap function $\Delta_k$ on hole kinetic energy $\varepsilon_k$[12], hence a finite gap slope b:

$$\Delta_k = a + b\varepsilon_k \qquad (2)$$

with a and b constants. Fig. 7 illustrates the consequences of Eq. (2): the minimum quasiparticle energy does not occur at the chemical potential $\mu$ but is displaced by the electron-hole symmetry-breaking term $\nu$. The coherence factors $u_k^2$ and $v_k^2$ are not 1/2 at this energy, which implies that quasiparticles are not charge neutral but positively charged on average. The tunneling asymmetry at low temperatures is given by $2\nu/\Delta_0$, with $\Delta_0$ the energy gap.

Hence the superconducting 'Fermi surface', defined as the locus in k-space of quasiparticle states of minimum energy, should be different (smaller) than the normal state Fermi surface, as sketched in Fig. 8. It is defined by $\varepsilon_k = \mu + \nu$, rather than $\varepsilon_k = \mu$. This should be observable in high resolution angle-resolved photoemission experiments, as shown schematically at the bottom of Fig. 8. In the normal state, the peak moves monotonically away from zero binding energy (dashed line) as k moves away from $k_F$; in the superconducting state, the peak moves first towards zero energy and only then moves away, at a slower rate[13]. The sharpest peak in the superconducting state, closest to zero energy, is at momentum $k_F'$, on the superconducting 'Fermi surface', rather than at the normal state value $k_F$. This measurement should also yield an experimental value of the fundamental parameter $\nu$. To our knowledge a careful analysis of photoemission data searching for this effect has not yet been performed, but resolution appears to be approaching the regime where it may be feasible[14]. We expect values of the parameter $\nu$ in the range 1 to 3 meV in high $T_c$ cuprates.

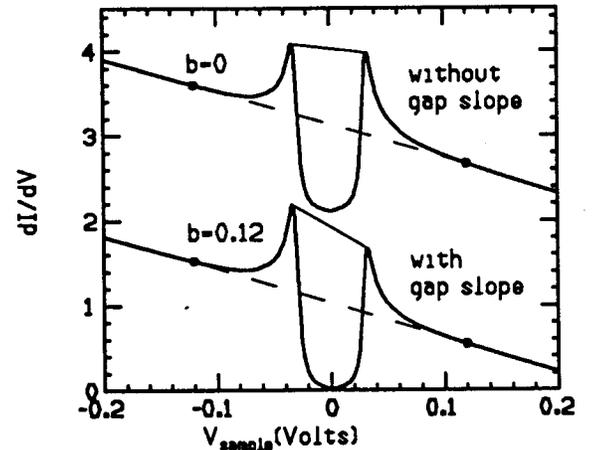

Fig. 6: Calculated NIS tunneling. The slope defined by the peaks is always smaller than that of the background if b=0.

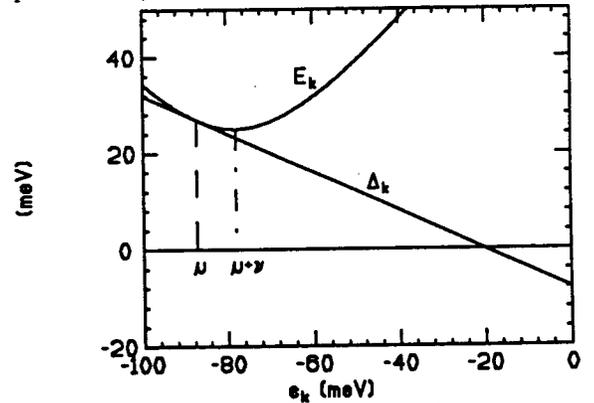

Fig. 7: Gap and quasiparticle energy versus band energy.

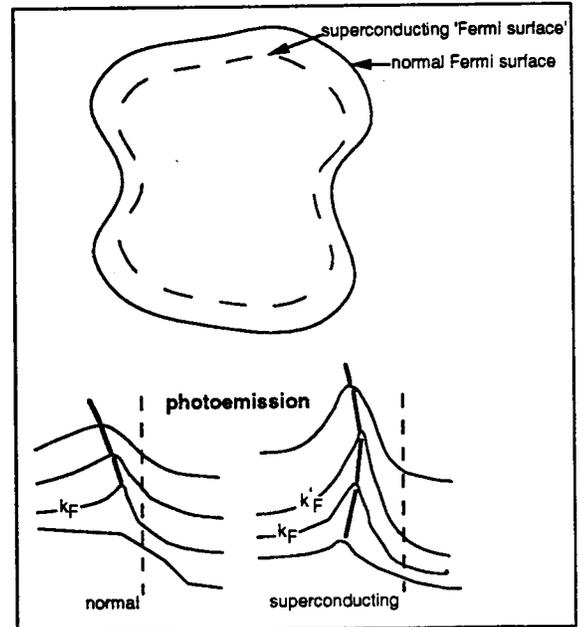

Fig.8: Schematic Fermi surface and photoemission spectra.

If a system does not have hole carriers it should not be superconducting according to the principles discussed in this paper. Hence we predicted in 1989, when electron-doped cuprate superconductors were discovered[15], that hole carriers had to exist in those materials[16]. Clear experimental evidence for hole carriers in the regime where the materials become superconducting was subsequently uncovered[17], confirming the prediction. More generally, there should be a strong correlation between the existence or nonexistence of superconductivity in a material and the sign of its Hall coefficient[18,19]. Figure 9 illustrates this correlation for the nonmagnetic elements. The only superconductors with negative Hall coefficient, La and Ga, presumably have both electron- and hole-like carriers at the Fermi energy.

Other predictions of the theory discussed here are at odds with the current consensus, namely: (1) that the superconducting state is s-wave for both electron and hole-doped cuprates, (2) that the carriers that drive superconductivity in the cuprates are holes in the planar oxygen $p\pi$ orbitals, (3) that the same basic principles discussed here that apply to the cuprates apply to <u>all</u> superconductors. The latter implies in particular that all superconductors have s-wave symmetry, have hole carriers, lower their kinetic energy when they become superconducting, and have sloped gap function, with slope of universal sign. We believe that future experiments and reinterpretation of the experimental results that led to the current consensus will demonstrate that all superconductors behave according to the fundamental principles of the theory discussed here. Because of the essential conflict with lattice stability brought about by the fact that antibonding states need to be occupied in superconductors, the search for higher $T_c$ materials will remain an elusive task.

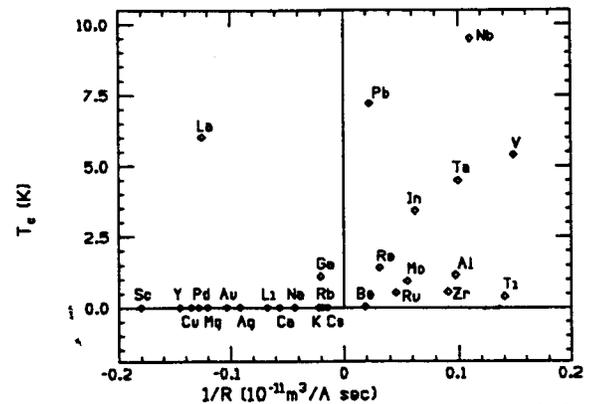

Fig. 9: Superconducting critical temperature of the elements plotted versus the inverse Hall coefficient at low temperatures and high fields. Data for the Hall coefficient were obtained from Ref. 20. Note that superconductors tend to have positive Hall coefficients.